\documentclass{article}
\usepackage{blindtext}
\usepackage[a4paper, total={6in, 8in}]{geometry}
\usepackage{natbib}
\setcitestyle{authoryear,open={(},close={)}} 

\usepackage{hyperref}

\usepackage{graphicx}%
\usepackage{multirow}%
\usepackage{amsmath,amssymb,amsfonts}%
\usepackage{amsthm}%
\usepackage{mathrsfs}%
\usepackage[title]{appendix}%
\usepackage{xcolor}%
\usepackage{textcomp}%
\usepackage{manyfoot}%
\usepackage{booktabs}%
\usepackage{algorithm}%
\usepackage{algorithmicx}%
\usepackage{algpseudocode}%
\usepackage{listings}%
\usepackage{comment}
\usepackage[acronym]{glossaries}
\usepackage{array}
\usepackage{booktabs}
\usepackage{authblk}

\usepackage{subcaption}
\usepackage{xcolor}

\providecommand{\keywords}[1]
{
  \small	
  \textbf{\textit{Keywords---}} #1
} 

\title {Am I Being Treated Fairly? \\[1ex] \large A Conceptual Framework for Individuals to Ascertain Fairness}

\author[1]{Juliett Suárez Ferreira}
\author[2]{Marija Slavkovik}
\author[3]{Jorge Casillas}
\affil[1]{Data Science and Computational Intelligence Institute (DaSCI), University of Granada.}
\affil[2]{Department of Information Science and Media Studies, University of Bergen}
\affil[3]{Data Science and Computational Intelligence Institute (DaSCI), Department of Computer Science and Artificial Intelligence (DCSAI), University of Granada.}

\date{}

\begin{document}

\maketitle

\begin{abstract}
 Current fairness metrics and mitigation techniques provide tools for practitioners to asses how non-discriminatory Automatic Decision Making (ADM) systems are. What if I, as an individual facing a decision taken by an ADM system, would like to know: \textit{Am I being treated fairly?}. We explore how to create the affordance for users to be able to ask this question of ADM.  In this paper, we argue for the reification of fairness not only as a property of ADM, but also as an epistemic right of an individual to acquire information about the decisions that affect them and use that information to contest and seek effective redress against those decisions, in case they are proven to be discriminatory. We examine key concepts from existing research not only in algorithmic fairness but also in explainable artificial intelligence, accountability, and contestability. Integrating notions from these domains, we propose a conceptual framework to ascertain fairness by combining different tools that empower the end-users of ADM systems. Our framework shifts the focus from technical solutions aimed at practitioners to mechanisms that enable individuals to understand, challenge, and verify the fairness of decisions, and also serves as a blueprint for organizations and policymakers, bridging the gap between technical requirements and practical, user-centered accountability.  
\end{abstract}


 \section{Introduction}
Artificial intelligence (AI) is increasingly used to automate aspects of operations in our society \citep{AIWatch2023}. The main motivation to use AI in operations, as with all automation, is to increase efficiency while reducing cost. Because the use of AI can have a direct and measurable impact on the lives of citizens\footnote{\href{https://automatingsociety.algorithmwatch.org/}{Examples} of how AI systems impact people's lives.}, we put a lot of focus on ensuring that AI is trustworthy \citep{lahusen_trust_2024} (i.e. lawful, ethical, and robust \citep{EU2019}). 

One of the more sensitive applications of AI is in its use as an aid in making decisions, namely as part of algorithmic decision-making (ADM). This is because the impact of such made decisions can be significant, for example, determining access to credit, employment, medical treatment, etc. \citep{EUPunderstanding_2019}. When ADM makes or influences important decisions about me, one concern that I have as a citizen is: Am I being treated fairly by this process? The field of AI ethics has been exploring how to achieve fairness, explainability, and accountability in various AI applications \citep{Huang2023}. But can all this work give an answer to the question: \textit{Am I being treated fairly by this AI system?}. 

There are different interpretations of fairness; but, the approach towards accomplishing fairness, however interpreted, needs to be both {\em substantive} and {\em procedural} \citep{EU2019}.  It should be noted that while the procedural dimension of fairness is associated with procedural fairness, it remains distinct from its definition, which is described as \emph{the process employed to reach or decide an outcome} \citep{Lionel2020}. A substantive fairness approach seeks to ensure an equal and just distribution of both benefits and costs, as well as to ensure that individuals and groups are free from unfair bias, discrimination, and stigmatization. A procedural fairness approach ensures the ability of a citizen to contest and seek effective redress against decisions made by AI systems and by the humans operating them. 

Answering the question \textit{Am I being treated fairly?} requires that substantive and procedural fairness be reified into an algorithmic process. Specifically, because a decision is now made by an algorithm, we need an algorithmic process to provide information about that decision. Digitalization increases discrimination risks, complicating institutional processes and making them less transparent. Moreover, algorithms process far more cases than humans, possibly causing unfair results by uncovering hidden patterns. Nonetheless, individuals have the right to question the fairness of decisions made about them. 

In this paper, we argue for the reification of fairness not (only) as a property of algorithmic decision making but as an epistemic right of an individual to attain information about decisions and use that information to contest and seek effective redress against those decisions. We refer to this epistemic right as {\em ascertainable fairness}.   

We examine key concepts, metrics, and methodologies from existing research not only in algorithmic fairness \citep{barocas-hardt-narayanan} but also in explainable AI \citep{BarredoArrieta202082}. \cite{EU2019} argue that in an ADM process, the entity responsible for the decision must be identifiable and that decision-making processes should be explicable. These are the two main characteristics for the ADM to be contestable, making a clear relationship between explanations and fairness. We analyze the role of contestability \citep{Lyons2021} for the procedural dimension of fairness,  the accountability field \citep{EU2019}, and recent advances in auditing machine ethics-based ADM algorithms. We observe that while much progress has been made towards empowering citizens to ascertain their own standing with respect to ADM fairness, ascertainable fairness is still not immediate.  

While traditional fairness mechanisms are designed mainly for developers and organizations to avoid discrimination in the ADM systems they develop and deploy, the approach in this paper is to shift the focus of algorithmic fairness away from tools solely intended for practitioners toward a suite of tools designed to empower citizens to actively validate, contest, and ensure their epistemic right to ascertain fairness in ADM systems.

After considering the advancements in the literature on AI ethics that contribute to attaining ascertainable fairness, we propose an ascertainable fairness conceptual framework that incorporates elements such as fairness of predictions, fairness of recourse, and mechanisms for contestation and request for audit. This framework is intended as a `blue-print' for policymakers and organizations that use ADM systems and those that develop ADM systems. It shows how people can be enabled to assess and challenge the fairness of AI decisions.

The scope of our proposal focuses on enabling individual users to actively engage and assess the fairness of decisions made by ADM systems, introducing tools that allow them to directly contest outcomes and seek redress, and access impartial third-party mediation when necessary. However, the framework is not without limitations. It assumes that users will have access to fairness metrics and explanations that may still be complex to interpret for non-expert users even with simplified tools. Moreover, the responsibility of developing contestation support mechanisms remains an open question. Additionally, even when it is designed to be broadly applicable across different domains where ADM systems are deployed, it may require adaptation to the regulatory environment of different sectors. 

The scope of our proposal focuses on enabling individual users to actively engage and assess the fairness of decisions made by ADM systems, introducing tools that allow them to directly contest outcomes, seek redress, and access impartial third-party mediation when necessary. However, the framework is not without limitations. It assumes that users will have access to fairness metrics and explanations that may still be complex to interpret for non-experts even with simplified tools. Moreover, the responsibility of developing some of its components remains an open question. Additionally, even when it is designed to be broadly applicable across different domains where ADM systems are deployed, it may require adaptation to different sectors and regulatory environments.
  
The structure of the paper is as follows. Section \ref{sec_2_background} establishes the foundational context on the substantive and procedural dimensions of fairness in ADM and explores the current advances in explainability, contestability, and accountability as part of the procedural dimension of fairness and their potential integration to ascertain fairness. Section \ref{sec_3_ADMfairness} examines the degree to which ascertainable fairness can be achieved through existing approaches to these dimensions. The paper continues in Section \ref{sec_5_ascfair} by outlining the process of inquiry and communication on fairness, together with the contestability dialog necessary to define ascertainable fairness. Section \ref{sec_6_framework} elaborates on the proposed conceptual framework for ascertainable fairness and the requirements that ADM systems must meet to enable it. Finally, we address the benefits and limitations of the study in Section \ref{sec_7_disc}, concluding with a synthesis of the results of our research in Section \ref{sec_8_conc}.

\section{Algorithmic Decision Making and Fairness Background}
\label{sec_2_background}

This section defines algorithmic decision-making (ADM) and reviews the state of the art on substantive and procedural fairness associated with ADM.  

\begin{figure*}
    \centering
    \includegraphics[width=0.99\linewidth]{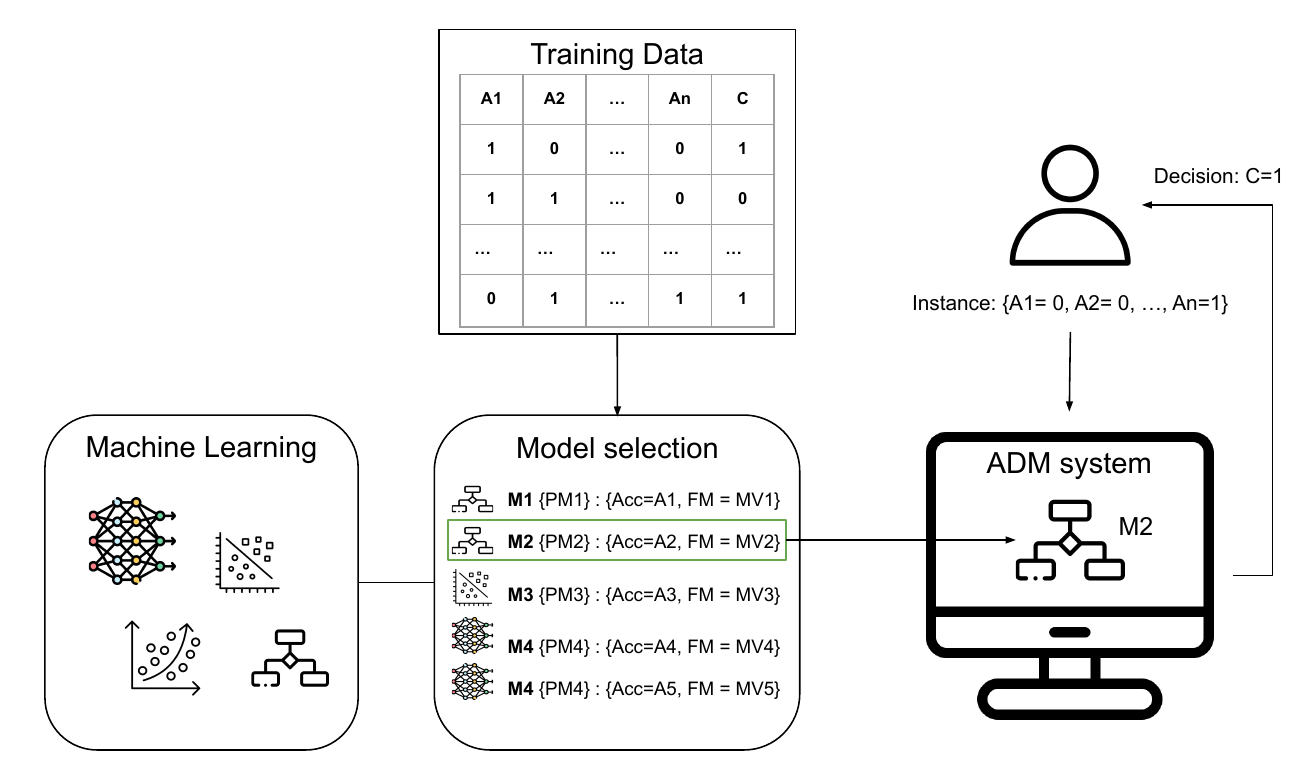}
    \caption{Selecting a model for an ADM system.}
    \label{fig:ADM}
\end{figure*}

ADM encompasses computational processes and systems for organizational decision-making \citep{EUPunderstanding_2019}. Human involvement in ADM varies from full automation (as defined by GDPR \citep{GDPR2016}) to assisted decisions, where end-users \textit{decision targets} are the recipients of decisions\citep{AYSOLMAZ2023}. We explore ADM systems delivering decisions directly to end-users, termed ``human out-of-the-loop'' \citep{Ivanov2023}. ADM relies on models for analyzing information and making predictions based on patterns \citep{Mohri2018}. These models guide decision-making and may include simpler techniques, such as linear regression or decision trees, or more complex techniques like neural networks. Figure \ref{fig:ADM} illustrates model selection and its central role in ADM. 

Regardless of the model type, end-users frequently perceive ADM systems as \textit{black boxes} due to the complexity and opacity of the underlying algorithms. This highlights the importance of transparency as a requirement for ADM systems, enabling users to recognize that they are interacting with such a system and to understand not only how decisions are made and why \citep{EU2019}, but also how to ensure that these decisions are fair. 

The definition of fairness encompasses two dimensions \citep{EU2019}; the substantive dimension of fairness focuses on ensuring equitable outcomes by addressing and mitigating biases in ADM systems, while the procedural dimension emphasizes transparency in decision-making processes, allowing users to understand, challenge, and seek redress for the decisions made by these systems. Together, these dimensions aim to promote fairness by ensuring both fair results and a fair process.

Although the substantive dimension of fairness is mainly used by practitioners to implement fairer ADM systems and it is also vital to address the quantifiable aspects of discrimination, such as biased data or algorithmic decisions; we consider that the procedural dimension of fairness may be investigated to support the epistemic right to ascertain fairness along with the information provided using the substantive dimension of fairness. We will analyze both dimensions of fairness, looking for tools available to the end-user to ascertain fairness.

\subsection{Substantive Dimension of Fairness: Algorithmic Fairness} 
\label{sbs:substantive_fairness}

The substantive dimension of fairness is defined as \textit{a commitment to ensuring equal and just distribution of benefits and costs ensuring that individuals and groups are free from unfair bias, discrimination, and stigmatization} \citep{EU2019}. 

The field of {\em algorithmic fairness} examines what makes ADM decisions fair constituting the implementation of the substantive dimension of fairness in ADM systems. Algorithmic fairness seeks to understand and correct the sources of unfairness \citep{pessach2022} identified as discrimination, resulting from human prejudice and stereotyping, and bias, arising from data collection and sampling \citep{Makhlouf2021}. This field offers \emph{metrics} for bias quantification and \emph{methods} to mitigate discrimination in algorithmic decisions, considering protected attributes such as race, sex, or age \footnote{Legally protected attributes defined in the \href{https://fra.europa.eu/en/eu-charter/article/21-non-discrimination}{EU Charter of Fundamental Rights. Title III: Equality. Article 21.}}. 

Several reviews classify fairness metrics into two main types: individual and group fairness \citep{Feuerriegel2020379, Makhlouf2021, Castelnovo2022}. \textit{Individual fairness} requires that similar individuals receive similar outcomes \citep{10.1145/2090236.2090255}. A common approach here is \textit{counterfactual fairness} \citep{DBLP:journals/corr/KusnerLRS17}, which holds that a decision is fair if it remains unchanged when an individual's sensitive attributes (e.g., race or gender) are hypothetically altered while all other factors remain constant. However, because counterfactual fairness depends on creating accurate and context-specific causal models \citep{DBLP:journals/corr/KusnerLRS17, kilbertus2017}, its application in different scenarios is often limited \citep{russell2017}.

\textit{Group fairness} demands that the model produce similar results for different groups defined by protected attributes. Common metrics used include Statistical Parity \citep{Calders2009}, Equalized Odds \citep{Hardt2016}, and Calibration \citep{Chouldechova2017}. However, even if a system satisfies group fairness, it may still produce individually unfair results. To bridge this gap, \emph{subgroup fairness} applies fairness constraints to both specific protected groups and finer subgroups, sometimes infinite, \citep{Mehrabi2021, kearns18a}.

A consensus has yet to be reached on the optimal metric for algorithmic fairness. Establishing these metrics involves not only mathematical, but also moral complexity \citep{Beigang2023}. As \cite{barocas-hardt-narayanan} explains, it is important to address these moral dilemmas in ADM fairness. Users may perceive unfairness if their moral values differ from those implemented by the ADM provider.

Another important area in the field of algorithmic fairness is the process of ameliorating the effect of bias on one or more protected attributes at different stages of the development of the ADM system (pre-, in-, and post-processing), called \textit{bias mitigation} \citep{barocas-hardt-narayanan, Bellamy}. The mitigation efforts themselves can introduce unfairness to an individual by increasing group fairness. This process remains undetectable to the end-user, who cannot ascertain, for instance, if the decisions provided have undergone alterations in a post-processing phase to meet certain fairness criteria, unless explicitly disclosed by the ADM system provider.  

From the definitions of substantive fairness and the work in algorithmic fairness, we can observe that the fairness of an ADM is seen as the responsibility of the practitioners developing the ADM systems and the organizations providing those systems. End-users are required to rely on organizations that develop ADM systems to have established the appropriate mechanisms to ensure fairness, or on independent institutions to certify that fairness requirements have been met \citep{Dowding2024}. 

The sociotechnical framework of ADM lacks mechanisms for users to independently verify the fairness of their treatment, leaving them reliant on trust in the system. This is not inherently problematic; for example, we trust the safety of prescribed drugs without personally verifying it, because these drugs meet set standards and the prescriber is qualified and accountable. However, ADM systems still lack established legal and regulatory oversight, making it crucial for users to independently assess such life-altering systems.

\subsection{Procedural Dimension of Fairness in ADM Systems}
\label{sbs:procedural_fairness}

According to \cite{EU2019}, the procedural dimension of fairness includes the ability to contest and seek redress against decisions made by AI systems and their human operators. For this to be effective, the responsible entity must be identifiable and the decision-making processes must be understandable. In this section, we delve into parts of this definition that extend the understanding of procedural fairness beyond the fair decision-making process studied by \citep{Decker2024} to include also mechanisms for redress and ensuring contestation. \cite{EU2019} definition comprises distinct aspects that merit separate analysis.

\begin{enumerate}
    \item Explicability of Decision-Making Processes: The decision-making processes of AI systems should be explicable, that is, transparent and understandable to the affected parties. Transparency is a fundamental aspect of trustworthiness, which will be examined through the lens of \textit{explainability}, given its significance to the end-user and consequent impact on fairness.
    \item Identifiability of Accountable Entities: For procedural fairness to be actionable, the entity responsible for the AI decision must be identifiable. This ensures that there is a clear line of accountability, making it possible to hold someone responsible for the ADM systems decisions.
    \item Redress: The ability to seek effective redress against AI decisions is fundamental to procedural fairness. This means that individuals or entities affected by AI decisions should have mechanisms to obtain remedies if the decisions are found to be unjust or erroneous.
    \item Contestation: The ability to dispute decisions made by ADM systems is crucial to ensuring procedural fairness. Users must have the chance to appeal and scrutinize these decisions, granting them the power to challenge the outcomes of such systems, and thus the possibility of an unfair treatment.
\end{enumerate}

Considering these elements of the procedural dimension of fairness, we will explore each of them giving an overview of the state-of-the-art across these areas that will help us settle the bases for our conceptual framework for ascertainable fairness.

\subsubsection{Explainability}

The field of explainability (XAI) \citep{BarredoArrieta202082} is crucial to achieve user-perceived fairness by clarifying decision processes to all stakeholders, showing rationale, and offering alternatives. It helps detect ADM biases by revealing decision-making attributes, logic, and organizational rules. Simply revealing the system isn't enough; the information must be made understandable with simple language, visuals, or interactive tools.

Current XAI efforts focus on explaining the model\footnote{The ADM system uses a pre-trained machine learning model, as shown in Figure \ref{fig:ADM}} used in the ADM system (the main characteristics of the model) or the results given by the model (the decisions of the ADM system; refer to \citep{BarredoArrieta202082, 9321372} for an overview of the methodologies). XAI methods explain the overall behavior of the model using global techniques, such as surrogate models \citep{Sharma2020166} or prototypes and criticism \citep{NIPS2016_5680522b}, or focus on individual predictions through local methods such as LIME \citep{ribeiro-etal-2016-trust}, SHAP \citep{NIPS2017_7062}, anchors \citep{Ribeiro_Singh_Guestrin_2018}, and especially and counterfactual\footnote{The term counterfactual is different in fairness and in XAI, counterfactual in algorithmic fairness implies causality.} explanations\footnote{Contrastive and counterfactual explanations are terms used interchangeably in the literature \citep{9321372}.}, which are valued for their clarity and actionability by suggesting minimal changes that could alter an outcome. 

Based on the analysis of existing methods for generating counterfactuals \citep{Laugel2023,verma2022counterfactual}, and supported by additional related articles, we analyze methods for generating counterfactuals with special attention to which of them made proposals for the evaluation of fairness. These works can be classified into three distinct categories based on their relation to fairness.. 

\begin{itemize}
    \item \textbf{Fairness of Predictions} (FOP), which evaluates the fairness of model predictions using generated counterfactuals and fairness metrics.
    \item \textbf{Fairness of Recourse} (FOR), which assesses the fairness of recourses (the actions the individual must take to change the decision) through generated counterfactuals and fairness metrics. These tools are significant because discrimination can go unnoticed: if an end-user faces more difficulty obtaining a different decision due to group affiliation or individual characteristics, they experience discrimination, even if the decision appears fair by standard metrics.
    \item \textbf{Fairness Assistance} (FA), which helps detect bias through visualization or linguistic output, serves as special cases of FOP without employing specific fairness metrics. 
\end{itemize}

A detailed classification of various contributions is presented in Table \ref{tab:counterfactualsmetrics} with a focus on the previous categorization, target stakeholders, fairness concepts considered, and whether they are model-agnostic. The relationship of the categorized works is not intended to be exhaustive; rather, it serves as the foundational basis for the body of literature that will inform our ascertainable fairness framework.

\begin{table*}[ht]
\centering
    \caption{Using counterfactuals explanations to check fairness. The table contains the Reference of each contribution, the classification we propose (Class) related to the fairness treatment, the stakeholder from whom is helpful the proposal (Helpful for), the fairness concept considered (FC: G(Group) and I(Individual)) and if the proposal is model agnostic (\checkmark) or not in the column MA.}
    \label{tab:counterfactualsmetrics}
    \begin{tabular}{llllll} 
    \toprule
    Reference & Class & Helpful for & FC & MA \\ 
    \midrule
    \cite{Sharma2020166} & FOP  & Developers & G & \checkmark \\
    \cite{Goethals2023} & FOP  & Policymakers & G &  \\
    \cite{10.1007/978-3-031-23618-1_27} & FOP  & Developers & G &   \\
    \cite{Galhotra2021} & FOP  & Developers & I\&G & \checkmark  \\
    \cite{Saloni} & FOP  & Developers & G &  \\
    \cite{Gupta2019} & FOR  & Developers & G & \checkmark \\ 
    \cite{Julius2020} & FOR & Developers & I\&G &   \\
    \cite{Yadav2021} & FOR & Developers & G & \checkmark \\
    \cite{Rawal2020} & FA & Decision Makers & G & \checkmark \\
    \cite{Cheng2020} & FA & Developers and users & I\&G & \checkmark \\
    \cite{Myers2020} & FA & Experts and nonexperts & G &  \\
    \bottomrule 
    \end{tabular}
\end{table*}

Counterfactual explanations clarify the decisions and minimal changes that end-users need to alter negative outcomes. They assess the fairness of ADM predictions and recourses using fairness metrics. In general, XAI helps end-users understand the logic of ADM, but fairness work is again more beneficial to practitioners than to users.

\subsubsection{Accountability}

Essential for transparency, trust, and fairness \citep{EU2019}, accountability involves different stakeholder roles in ADM systems and is crucial to procedural fairness. As it requires stakeholders to assume responsibility by providing justifications and ensuring transparent system development and deployment \citep{EU2019, Reuben2018, Horneber2023}. A proposed solution is public reason, where AI providers must normatively justify their systems to gain societal trust \citep{Binns2018}. 

Key mechanisms of accountability include \textit{auditability}, independent evaluations of \textit{black box} systems to confirm fairness \citep{Tang2023, toreini_fairness_2024} and \textit{redress}, which offers users a way to change unfair decisions through algorithmic recourse \citep{natalia2023, Karimi2022a, Karimi2022}. 

Audiatability mechanisms help users verify fairness, but must be tailored to their needs; otherwise, trust in independent auditors is essential \citep{Dowding2024}. In disputes, auditing offers impartial conflict resolution. Redress, on the other hand, upholds fairness by compensating users' non-discrimination rights \footnote{\href{https://fra.europa.eu/en/eu-charter/article/21-non-discrimination}{EU Charter of Fundamental Rights. Title III: Equality. Article 21.}}.

\subsubsection{Contestability}

A contestation process involves providing clear pathways for individuals to question and dispute automated decisions. Although contestability is not considered a principle for trustworthy AI by \cite{EU2019}, it is considered a pathway to fairness. Its role in AI ethics is debated, drawing attention from researchers. Some scholars regard contestability as a post-hoc tool to challenge decisions \citep{Lyons2021}, while others see it as a design feature \citep{lyons2021fair, Almada2019}. We contend that both elements are crucial. ADM systems should be inherently contestable, and a post-deployment mechanism should allow users to exercise their GDPR-guaranteed right to contest \citep{GDPR2016}. Contestability should let users challenge decisions, but also fairness metrics, involved attributes, and their impact on outcomes. 

Although certain studies indicate that end-users may lack the knowledge required to effectively challenge an ADM system \citep{Alfrink2022, Lyons2021}, we consider contestation mechanisms to be a means of empowering and conferring upon them the right to contest, as articulated in other works \citep{Lyons2021, Henin, Vaccaro2020}.

\cite{leofante2024contestable} analyzes contestability and supports using computational argumentation to achieve it. The authors outline that ADM systems need an explanation method and a redress method to facilitate contestation. They also propose \textit{ground generator method}, enabling automatic generation of contestation grounds. Both components should interact, allowing for a conversational contestation process.

To ascertain fairness, the end-user will have the role of contester and the ADM system is the contested entity. The ground generator method can be useful for end-users that are not experts in the domain as a support tool that gives them the foundations for their contestation. We argue that contestability, as a mechanism post-decision, has the potential we need for challenging the ADM in terms of fairness, not only the fairness of predictions or the recourses, but also the inputs used, the fairness metrics taken into account, and the process of decision making. 

From the end-user's point of view, interacting with an ADM system designed to allow the contestation of its decisions is highly valuable \citep{Yurrita2023}; the process will involve an exchange of evidence that makes contestability suitable for use for the computational implementation of the procedural dimension of fairness as a complement to explainability. 

\section{Algorithmic Decision Making and Ascertainable Fairness}
\label{sec_3_ADMfairness}

In this section, we discuss the extent to which ascertainable fairness can be achieved by existing approaches to substantive and procedural fairness.

\subsection{Algorithmic Fairness and Ascertainable Fairness}
\label{sbs:substantive_ascertainable}

For the purposes of ascertainable fairness,  unfairness should be evaluated against a subset of personal characteristics that encapsulates the group (collective) identity of the end-user, and this subset of characteristics can vary between individuals.  A \textit{collective identity} is one that is shared with a group of others who have (or are  believed  to  have)  some  characteristic(s) in common \citep{Ashmore2004}. Beyond the legally defined set of protected attributes, individuals can identify with different groups at the same time and identify more strongly with some of these groups over others. Individuals try to find balance in their need to belong and in their need to be different \citep{Hornsey2004}.

\begin{figure*}[ht]
\centering
\begin{subfigure}[t]{.49\textwidth}
    \centering
    \includegraphics[width=\linewidth]{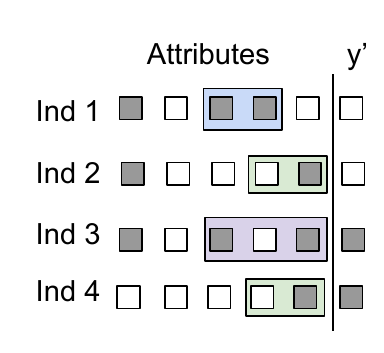}
    \caption{Individual perceptions of fairness can vary, as each person may possess a unique set of attributes that contribute to their collective identity.}
    \label{IndP}
\end{subfigure}
\hfill                    
\begin{subfigure}[t]{.49\textwidth}
    \centering
    \includegraphics[width=\linewidth]{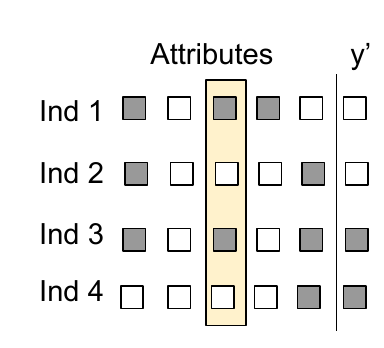}
    \caption{The implementation of fairness within ADM systems typically involves organizations adopting a particular concept of fairness that considers specific protected attributes.} 
    \label{OrgP}
\end{subfigure}
\caption{Variations in the perception of fairness by individuals and organizations.}
\label{fig:FairnessPerception}
\end{figure*}

An organization may prioritize fairness for one protected group over others. However, the attributes they use to demonstrate system unfairness may differ from those reflecting the collective identity of individuals using the ADM system. This need for balance is evident in aligning with the protected group each individual identifies with. Figure \ref{fig:FairnessPerception} illustrates differences in individual perceptions of fairness versus fairness implementation in ADM systems.

Mitigation techniques and fairness metrics help practitioners create models for ADM systems that align with non-discrimination standards. However, these are not directly available tools for individuals to evaluate fairness, as they often lack expertise and access to necessary data. Many algorithms are either non-transparent or proprietary, restricting public scrutiny. This limits individuals' ability to assess the fairness of algorithms affecting their lives. Fairness metrics can inform users, but understanding them is required for individual and societal benefits \citep{Nieminen2024}. 

\subsection{Procedural Fairness and Ascertainable Fairness}
\label{sbs:procedural_ascertainable}

We argue that the procedural dimension of fairness gives more resources to end-users, which allows them to ascertain fairness providing not just a metric, but also the understanding of how decisions were taken (explainability), the possibility to challenge them (contestability) and obtain effective compensation (accountability and redress), which supports the epistemic right of ascertainable fairness. However, these fields are not put together in the service of procedural fairness. Furthermore, although comparatively much work has been done in the field of explainable AI, the same effort is not matched in contestability, the technical aspects of accountability such as auditability and redress. The existing work in these fields studied in the previous section provide the building blocks for our ascertainable fairness framework. 

\section{Ascertainable Fairness}
\label{sec_5_ascfair}

\begin{figure*}[ht]
\centering
\begin{subfigure}[t]{.49\textwidth}
    \centering
    \includegraphics[width=\linewidth]{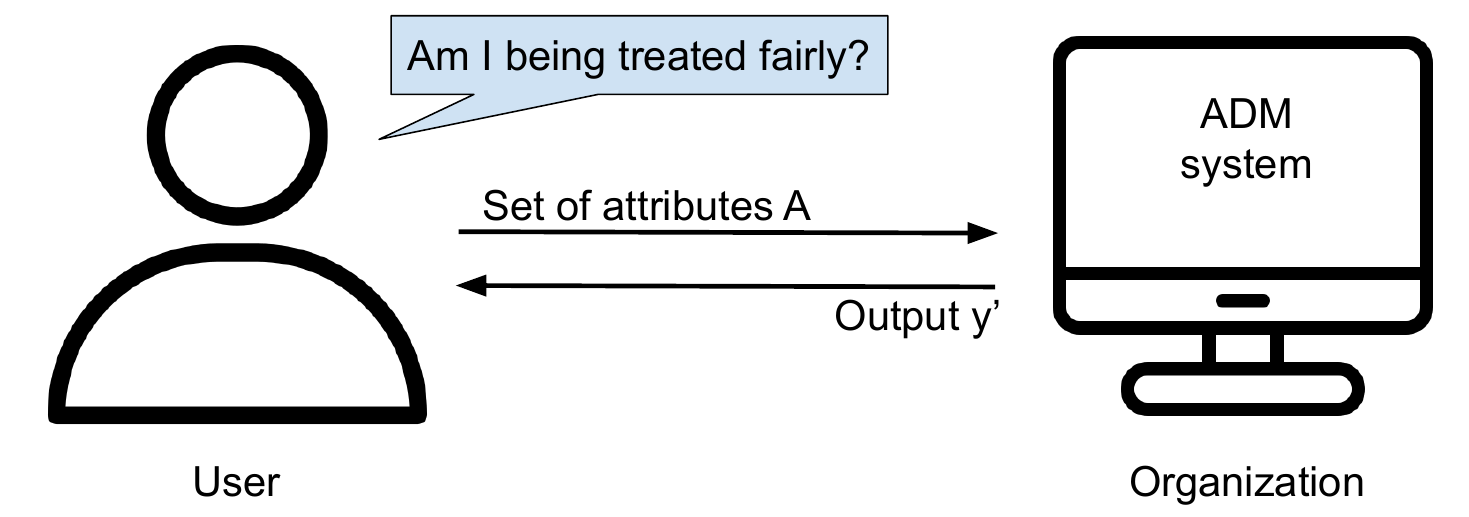}
    \caption{Current state: individuals lacks of means to ask and be informed about fairness}
    \label{AF_question}
\end{subfigure}%
\hfill                      
\begin{subfigure}[t]{.49\textwidth}
    \centering
    \includegraphics[width=\linewidth]{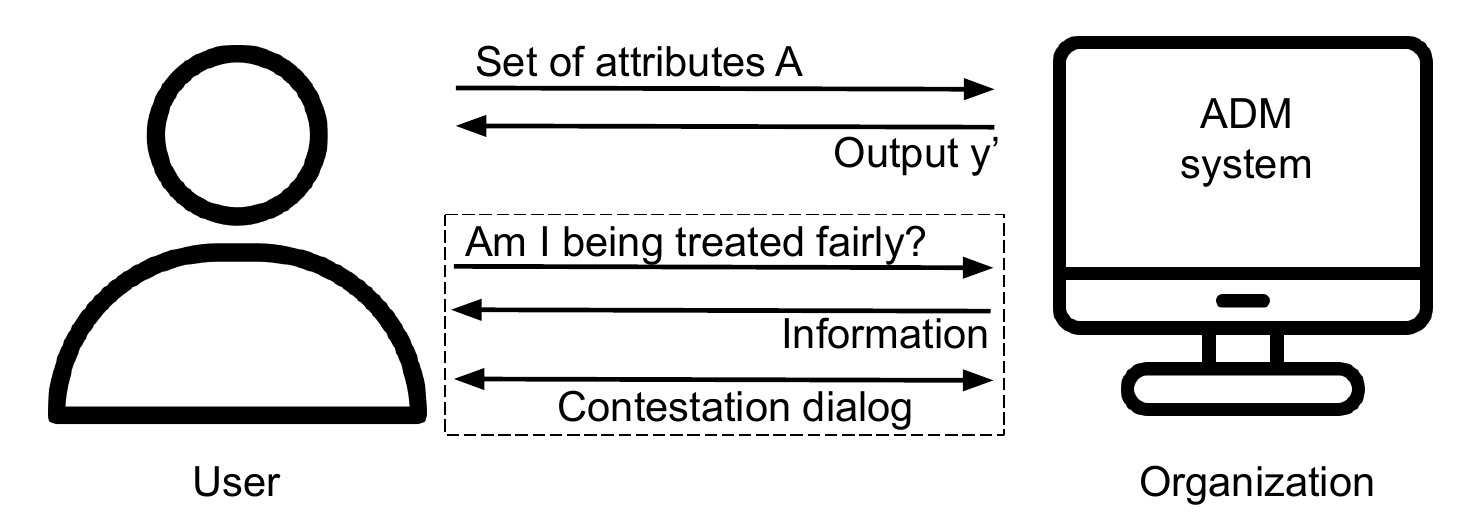}
    \caption{Ascertainable Fairness: individuals can ask and be informed about fairness} 
    \label{AF_framework}
\end{subfigure}
\caption{Ascertainable Fairness: from an unaddressed question to it's operationalization in ADM systems.}
\label{fig:AF}
\end{figure*}

In this section, we set the prerequisites for a framework that provides individuals with the possibility to ascertain the fairness of decisions made by ADM systems and provide our definition of ascertainable fairness. In the current landscape, individuals lack a tool that allows them to ascertain the fairness of the decision taken by the ADM system (Figure \ref{AF_question}). 

In simple terms, we consider a framework built around an ADM system that provides direct users with information about the decision and an accountable channel to contest that decision (engage in a contestation dialogue), as illustrated in Figure~\ref{AF_framework}.

What does it mean to allow an individual user to ask \textit{Am I being treated fairly?} and to provide information to answer that question meaningfully and the mechanism for challenging the ADM system through a contesting dialog are explored in the subsequent sections.

\subsection{Asking and Informing About Fairness}
\label{subsec:ask_inform}

Fairness can be evaluated at both the collective and individual levels. To ensure fairness, individuals must be able to assess whether their characteristics have resulted in biased decisions. Assessing whether actions to change a decision are feasible is key to understanding potential unfairness. We argue that contestability mechanisms should uncover sources of discriminatory practices. Ultimately, individuals have the right to receive explanations \citep{GDPR2016} of the decisions made by ADM systems that are helpful to determine whether they are warranted or if they stem from biased or discriminatory methods.

Consider a scenario in which an ADM system $S$, developed or deployed by some organization, takes input a data point, a set of known attributes $A = {a_1, a_2, .. a_n}$, whose values describe a particular problem instance associated with an individual.  $S$  produces on output $y'$, which is the decision taken in our case (See Figure \ref{AF_question}). The individual should be able to inquire about and determine how fair the decision made by $S$ is in their case. We say that {\bf the end-user can ascertain the fairness of the decisions that affect them}.

A person can be treated unfairly because they are members of a group that is treated unfairly. When we compare groups against groups, we evaluate fairness on a collective level. In this case, fairness is evaluated against a subset of attributes denoted as $P_A \in A$, whose values encapsulate the group (or collective) identity of an individual. When we consider specifically whether a decision is discriminatory, $P_A$ is called ``a protected'' set of attributes. An individual may experience unfair treatment as a member of a protected group, even when the outcome distribution between groups appears equitable. 

The collective identity of a person is unique; the attributes that are considered important to an individual may not be the attributes considered by the organizations that provide the ADM systems, as we have seen in Figure \ref{fig:FairnessPerception}. Moreover, the perception of fairness is individual and can be manifested differently in different end-users. A user may attempt to find out if particular attribute values are causing the different treatment they are facing, irrespective of their group affiliation. Thus, to enable users to ascertain fairness, they should be able to verify the satisfaction of fairness of the decision they have been subject to with respect to the concept of fairness with which they identify (i.e., different metrics, different combinations of attributes), and also the actions they need to take to change it.

As we have seen, the XAI field already has tools to implement some part of the procedural dimension of fairness along with the metrics derived from the field of algorithmic fairness. These tools have been designed to verify the fairness of predictions made by the ADM systems (FOP) and fairness of recourses (FOR) (i.e. the actions that the final users of the ADM systems need to do to change the decision). Nevertheless, the result of these verification processes is a value of a fairness measure, but is a numerical value from a measure or a simple true/false derived from them enough? We consider it insufficient and hard to understand but helpful as grounds for establishing a contestation dialog (as suggested by \cite{leofante2024contestable}) with the ADM that should justify the relevance/suitability of the metrics, attributes and processes used to make the decision.

Informing about fairness requires mechanisms that allow users to verify the absence of bias and discrimination. Bias can arise from the data that affect the algorithm (Data to Algorithm), bias originated by the algorithm used (Algorithm to User), and biases in users might be reflected in the data they generate (User to Data) \citep{Mehrabi2021}; we consider that having transparency on how organizations handle bias risk is the way that users can have to uncover unfairness. Discrimination may arise from human involvement or organizational policies; contestability mechanisms should help uncover discriminatory practices within organizations.

In addition, when an individual perceived an unfair decision, the sole method for its acceptance is to receive a justification of the reasons behind it. \textit{Is it justified?} will depend on the specific problem. In algorithmic fairness, sometimes different thresholds are necessary for different groups, and this can cause discrimination depending on the problem. Examples: consider the case of recidivism prediction algorithms; an African American may receive higher risk scores, a form of discrimination that is not justified, as there is no evidence to suggest that one's ethnicity makes them more likely to reoffend. Research has shown that Indian-Americans are more prone to diabetes, leading to different base rates of risk between groups \citep{10.2337/cd15-0048}. In this case, differential treatment could be considered a justified discrimination based on empirical evidence. In other cases, positive discrimination is imposed to achieve, for example, gender equality in education due to the historical bias that women and girls have suffered\footnote{UNESCO \href{https://www.unesco.org/en/gender-equality/education}{strategy} for gender equality in education.}. 

Determining if the treatment is justified involves justifications, not explanations. Therefore, justifications are essential so that individuals can ascertain the fairness of the decisions that affect them. Justifications provide context to particular cases where explanations might seem biased, offering insight into legitimate internal or external norms. So, if an individual has been treated unfairly, is there a justification that the ADM system $S$ can provide? Justifications can be based on external or internal rules (norms) of organizations based on requirements that are generally outside the algorithmic system, which, as mentioned in \citep{Reuben2018, EUPunderstanding_2019}, applies to requirements for ADM and is crucial for accountability.

In summary, asking and informing about fairness involves enabling individuals to assess whether ADM systems treat them fairly at both individual and group levels, understanding the feasibility of changing decisions by providing mechanisms to uncover discriminatory practices, and offering justifications (not just explanations) for decisions, since fairness perceptions vary among users and can arise from multiple sources including data bias, algorithmic bias, and organizational rules.

\subsection{Contesting Dialogue}
\label{subsec:dialogue}

\begin{figure*}
    \centering
    \includegraphics[width=0.99\linewidth]{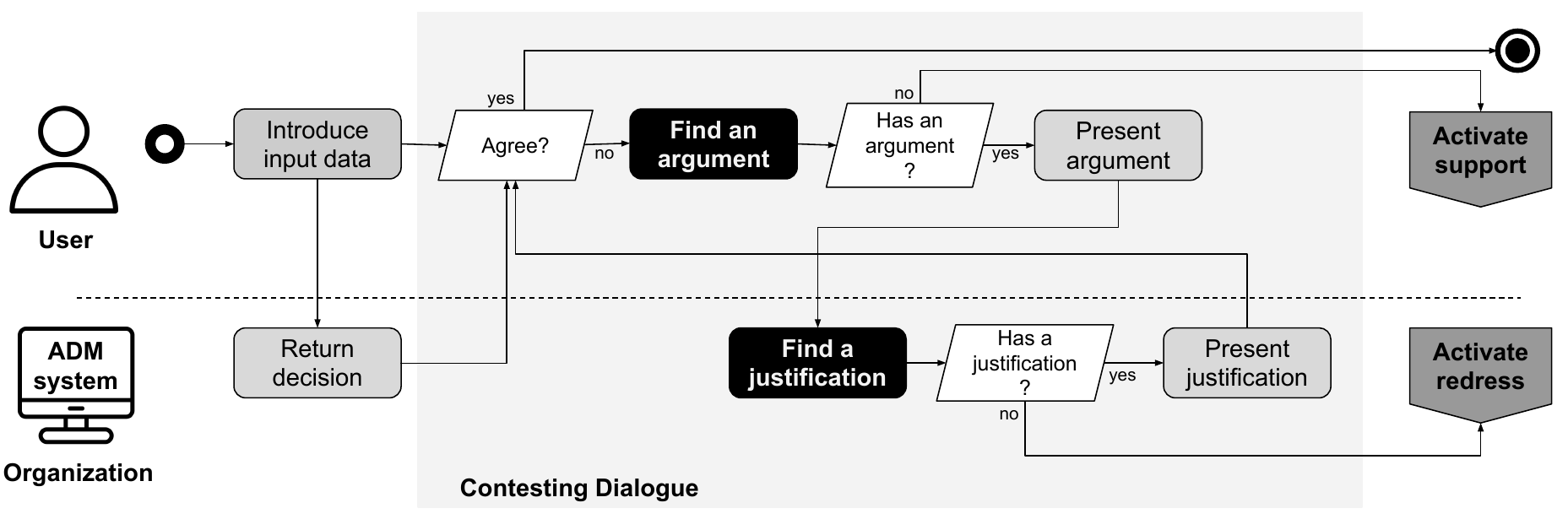}
    \caption{Contesting dialog}
    \label{fig:contesting-dialogue}
\end{figure*}

We now consider what allowing for a contesting dialogue can mean. \cite{leofante2024contestable} discusses how to build a computationally plausible contestation process based on argumentation. We consider that this general process of contestations based on the exchange of arguments operationalizes contestability as an essential mechanism of the procedural dimension of fairness to allow users to ascertain fairness (see Figure \ref{fig:contesting-dialogue}). 

End-users or other stakeholders can challenge the suitability of the ADM system for a task. Apart from contesting the system itself concerning its adequacy and appropriateness for addressing the problem at hand, we provide a list of fairness-related contestations they can raise:

\begin{itemize}
    \item The output of the ADM system. The user could disagree with the decision received, this is the primary contestation.
    \item The use (or non-use) of attributes. This should include the particular combination of attributes that the user is identifying with.
    \item The importance of an attribute or the correlation of an attribute with the output received.
    \item The fairness measure used. This will challenge the concept of discrimination utilized by the organization that supplies the ADM system.
    \item The fairness of the predictions, explanations, and recourses received by the end-user as well as their validity.
    \item The validity (legitimacy) of the justifications given in the contestation process.
    \item The variation of the outcome for a different case similar to the end-user's case.
    \item An error in the applied norms/rules specific to the solution .
    \item An internal rule of the organization revealed in a justification.
\end{itemize}

The result of a contestation process to ascertain fairness should be a legitimate justification that convinces the individual of the fairness of the treatment received; otherwise, if discrimination is exposed, this process could potentially lead to a change in the decision received using a redress mechanism. If contestation shows discrimination or the user doubts the justification's legitimacy, they should be able to request an audit from a relevant regulatory authority.

\subsection{Ascertainable Fairness: a Definition}

After analyzing different aspects of substantive and procedural dimensions of fairness as well as the interoperability of different fields in response to the epistemic right of an individual to ascertain the fairness of the ADM systems decisions that affect them; we can describe ascertainable fairness as \textit{ the ability of end-users of ADM systems to authenticate the concept of fairness with which they resonate, considering their shared identity and individual traits, potentially identifying discrimination sources, and acquiring justifications via a contestability mechanism, leading to either verification of fairness or availability of redress and / or audit results}.

\section{Ascertainable Fairness: a Conceptual Framework}
\label{sec_6_framework}
 
\begin{figure*}[h!]
    \centering
    \includegraphics[width=0.85\linewidth] {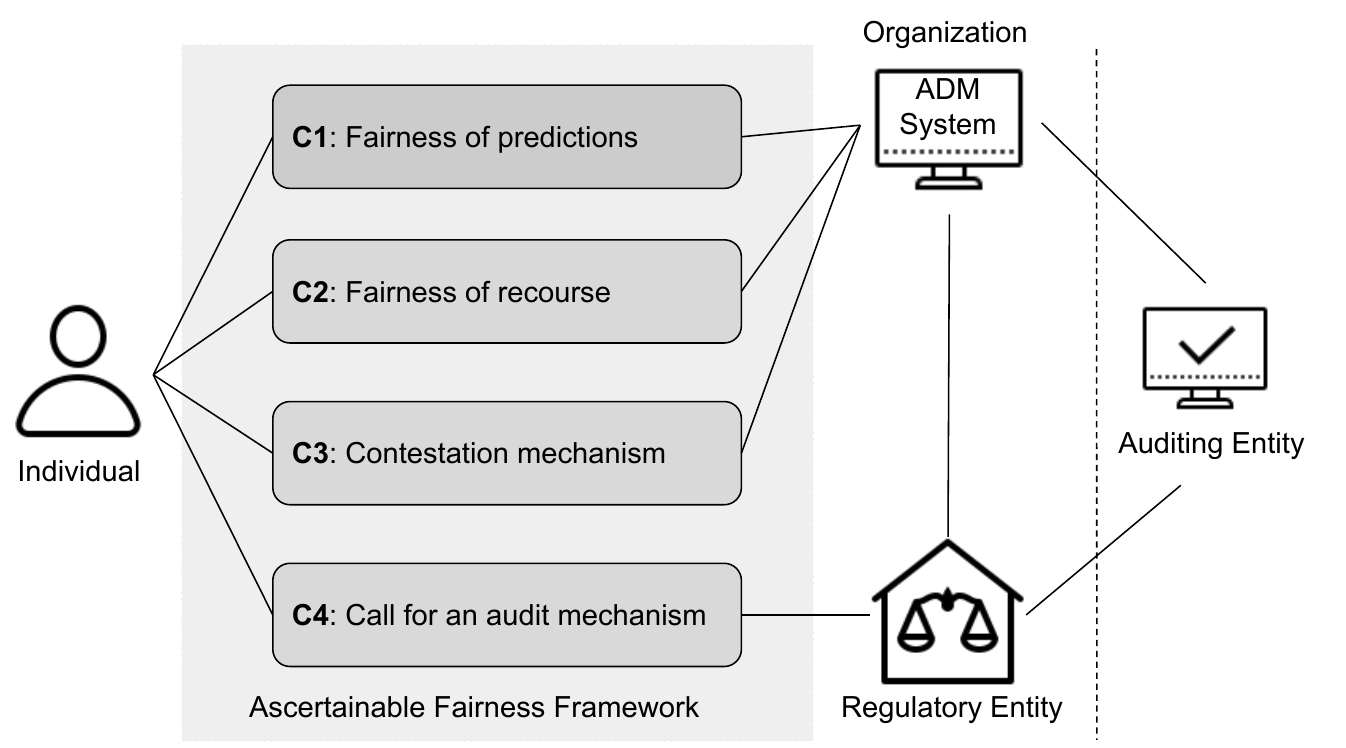}
    \caption{Ascertainable Fairness Framework. The end-user interacts with different tools to ascertain the fairness of the decision received.} 
    \label{fig:VF}
\end{figure*}

In this section, we present a new conceptual framework that will support end-users in the process of ascertaining fairness by providing different components illustrated in Figure \ref{fig:VF}. 

We propose the combination of different tools to ascertain fairness with the following components:

\begin{itemize}
    \item C1. A tool for checking the fairness of predictions. The output output $y'$ received by the user giving the attributes provided. The tool for assessing the fairness of the predictions will have access to query the system $S$ and build a parallel model that can simulate the behavior of the system and verify bias in the algorithm decision.
    \item C2. A tool for checking the fairness of recourses. The explanations $E$ given to the user may point to changes that the user can apply to change the final outcome. Are those actions i.e. recourses fair for the user? The tool to check the fairness of the recourses will use the system $S$ and the explanations $E$ given to the user to check if the changes the user needs to make to receive a positive decision are fair. 
    \item C3. A contestation mechanism that allows the user to challenge the ADM. This tool will use the results of components 1 and 2 as well as the explanations $E$ provided by the ADM and possibly a ground generation tool to establish an exchange of arguments with the system that need to provide justification to the individuals not just for the decision made but also for the process to obtain it, as well as the different elements that can be subject to a contestation. 
    \item C4. A mechanism for reporting the organization to a regulatory entity and requesting an audit. If there is a conflict between the organization and the end-user, this channel serves as the end-user's final option, not to ascertain fairness, as it is already confirmed for the user, but to request validation of the decision from a regulatory entity and ultimately seek redress. 
\end{itemize}

Components 1 and 2 integrate elements from both the substantive and procedural aspects of fairness. In contrast, component 3 employs the contestability element of the procedural dimension, utilizing the outputs of components 1 and 2 along with an internal mechanism for ground generation. Component 4 enables users to seek external help if discrimination is suspected and also falls within the procedural dimension of fairness. Figure \ref{fig:VF} illustrates how these components interact with the stakeholders we have identified. The fairness metrics adopted by components 1 and 2 should be able to verify the metrics reported by the organization and other metrics, as well as various combinations of attributes, thus allowing beneficiaries to confirm their self-identity.

Figure \ref{fig:VF} illustrates how these components interact with the stakeholders we have identified. \textit{End-users} are the individuals affected by the decision of the ADM system. The end-users are responsible for questioning the decision to which they were subject and taking the necessary steps to challenge the systems and possibly reverse the decision. They should define the set of attributes $P_A \in A$ that correspond to the group with which they are identified and use a framework to verify whether it is discriminated against, taking into account this identity.

\textit{Organizations} are the responsible o developing the ADM systems. Within organizations, the \textit{ practitioners} are members of organizations with different roles in the development and deployment of ADM systems. They are are not represented in the figure but are worth to mention for their role in the developing and deploying process. Organizations and practitioners are responsible for the implementation of all the mechanisms to avoid unfairness in the development of ADM systems and to disclose the fairness definitions used to evaluate the proposed solution. Organizations provide a mechanism to challenge their system; this mechanism should be different from the redress mechanism or may include it.

\textit{Regulatory Entities} are competent authorities responsible for ensuring nondiscriminatory ADM systems. Regulatory entities should create mechanisms for appealing a decision to provide the user with tools, external to the organizations, that can lead to an audit of the process by an auditing entity.

In addition to the components that will help the user verify the treatment received, Figure \ref{fig:VF} shows an additional entity that could act on behalf of regulatory entities in two main forms: (1) an audit process made by an independent entity in the form of a certification that checks a specific requirement, fairness in this case; (2) an audit process triggered by a reclamation related to unfairness originated by the user. \textit{Auditing Entities} are independent entities that perform conformity assessments in ADM systems. The auditing entities should be designated by these regulatory entities. 

To ensure ascertainable fairness, an ADM system should meet certain requirements. It must allow unlimited petitions with new data input, providing decisions to facilitate fairness verification. The system should disclose all attributes used in decision making, ensuring transparency between user input and model features. It must provide users with explanations and recourses (minimal changes needed to alter decisions) enabling contestation. A redress mechanism must be in place to compensate affected users and implement corrective actions to prevent recurring unfair outcomes. Finally, the system must disclose its fairness criteria and report corresponding values, helping stakeholders to assess its fairness approach. These requirements collectively empower users to verify, contest, and seek redress in ADM decisions.

In the subsequent subsections, we will outline the various elements of the framework, discussing both the existing work applicable to each component and the gaps that need to be addressed.

\subsection{Components 1 and 2: Fairness of Predictions and Fairness of Recourses}

The field of algorithmic fairness offers metrics that organizations can use internally to guarantee fairness; however, the issue is that fairness is an individual experience and the perception of the implemented fairness concept can differ among people, as well as the attributes that individuals view as part of their collective and personal identity.  The components 1 and 2 of the proposed conceptual framework aim to address this problem. Providing the end-user with two components that combine different fairness metrics and the possibility of defining the subset of attributes they consider their personal identity allows them to express their perception of fairness accordingly. The XAI field is crucial for components 1 and 2 since explanations have been extensively used to check the fairness of predictions and fairness of recourses.

Table \ref{tab:counterfactualsmetrics} lists some tools developed to assess the fairness of predictions and the fairness of recourses, as well as those aiding visual fairness understanding. They function by using the user's identified attributes to apply fairness metrics to predictions or recourses, evaluating various fairness concepts.

The open area is to make these tools useful not only for practitioners, but also to end-users. Even when these tools can be used to verify fairness the limitations are clear: 1) there is no unified metric to quantify unfairness, and 2) the definitions have incompatibilities with each other. Therefore, the organization's definition of ADM fairness might not align with the end-user's perception, as the attributes or metrics deemed relevant by the organization may differ from those considered important by the end-user.

We also consider that it is unclear which stakeholder will be responsible for building components 1 and 2. They should be created outside the organizations, probably by some body designated inside the regulatory entities that can create standard tools to check the fairness of prediction and fairness of recourse the same way they will make possible the creation of sandboxes for organizations to develop, train, validate, and test their ADM systems. 

Components 1 and 2 are not enough to operationalize the procedural dimension of fairness; nevertheless, these tools clarify the interpretation of fairness of the final user by providing an adjusted measurement according to a different point of view and their personal characteristics, and their results can serve as grounds for the contestability component. 

\subsection{Component 3: Contestation Mechanism}

Contestability is an instrument to be used to answer questions about fairness treatment that may arise when an individual receives an automatic decision. By challenging the ADM, the end-user will look at the treatment received as an individual or as a member of a particular group. This field gave us the guidelines for the design of component 3. We also argue that this component can use components 1 and 2 to create arguments to challenge the ADM system. 

We have enumerated potential contestations that the user can raise. Component 3 will grant the user access to a procedure where the system's decision can be contested and possibly altered. Moreover, this tool address different concerns by challenging the system based on the inclusion or exclusion of attributes, examining the potential misuse or lack of attributes, and their correlations with the decision. It will clarify the reasoning behind the choice of a specific fairness measure. The contestation tool should disclose the presence of internal rules that influence the decision, such as random choices, post-processing methods, or business rules that can include positive discrimination. The final arguments of the ADM system should be justifications, not explanations, illustrating why the decision is fixed and unchangeable.

The contesting dialog, visually represented in Figure \ref{fig:contesting-dialogue}, serves to identify two parts of the process that need further exploration: the method used by the user to formulate an argument and the approach utilized by the system to derive a justification. 

The authors \cite{leofante2024contestable} sheds light on the development of a computational framework for contestations through argumentation, utilizing a ground generator tool to help users throughout the procedure. However, we acknowledge that considerable advancements remain necessary in this domain; it is clear that organizations must provide contestation mechanisms that allow the user to challenge the ADM system, but it is not clear who is responsible for creating the ground generator tool to assist the end-users.

The final resolution is either a justification that the end-user will accept as valid, a possible demonstration of an error that could trigger the redress mechanism $R$ of the ADM system to change the decision and provide compensation (and the subsequent action within the organization to prevent similar cases), or a disagreement between parts. The final two options can lead the end-user to request an audit to a relevant regulatory entity. 

\subsection{Component 4: Call for an Audit Mechanism}

An established mechanism for reporting organizations upon substantiation of discrimination should be accessible to end-users to facilitate audit requests. Similarly, any internal rules that result in discriminatory practices, once identified, ought to be reported. Nonetheless, mechanisms for reporting organizations that implement inequitable ADM systems remain unestablished or inadequately delineated, aside from conventional legal channels, which may be ambiguous to end-users of ADM systems.

The accountability field had helped us to specify the responsibilities of different stakeholders interacting with ADM systems; moreover, it defines auditability as one of their main aspects, which is important in component 4 since it is the process that is defined and will start after the user reports an unfair treatment. The audit process, though outside the ascertainable fairness framework, is essential as it allows regulatory bodies to mediate conflicts between organizations and individuals or to provide organizations with a means to demonstrate compliance and build trust.

The Artificial Intelligence Act\footnote{AI act \href{https://eur-lex.europa.eu/eli/reg/2024/1689}{text}.} contains some legal terms for what we call regulatory entities that can be extrapolated to our context. A \textit{notifying authority} means the national authority responsible for setting up and carrying out the necessary procedures for the assessments, designation and notification of \textit{conformity assessment bodies} and for their monitoring. While a \textit{conformity assessment body} means a body that performs third-party conformity assessment activities, including testing, certification, and inspection. We argue that, no matter the risk of a system, the mechanism to notify issues related to discrimination should be created. These predefined notifying authorities must ensure the assignment of a conformity assessment body to verify the particular ADM. The creation or clarifications of such mechanisms, we argue, are important to increase the trust in ADM systems; the end-users shall be sure that if something is wrong, the mechanism for appeal to call for an appeal to a regulatory entity is available and clear.

\section{Discussion}
\label{sec_7_disc}

In this section, we examine our theoretical and practical contributions as well as the open areas and limitations of our approach. The proposed framework integrates fairness, explainability, contestability, and accountability into four components that enable users to ascertain fairness.

\subsection{Theoretical and Practical Contributions}

This paper advances the theoretical understanding and practical implementation of fairness in ADM systems. We distinguish between theoretical contributions that expand current concepts and practical contributions that provide implementable solutions.

As theoretical contributions, we introduce ascertainable fairness, treating fairness as an individual's right to access and verify decision-related information. We integrate fields such as algorithmic fairness, explainable AI, contestability, and accountability into a conceptual framework that expands the understanding of procedural fairness and provides a foundation for fairness verification mechanisms. Finally, we propose a user-centered fairness perspective, linking individual perceptions to collective identity, enabling fairness authentication based on personal context.

Our research offers two practical contributions. We propose four components combined in a framework for ascertainable fairness: a tool to verify prediction fairness, a mechanism to assess fairness of recourse, a contestation method, and an audit mechanism. These components are accompanied by specific requirements that ADM systems must meet to enable ascertainable fairness in practice and the necessary interactions between different stakeholders in the system. Moreover, we provide customized guidance for stakeholders: processes for end-users to verify and contest decisions, strategies for organizations to implement fairness and contestability, requirements for practitioners to build fairer systems, and guidance for regulators on auditing fairness claims. 

\subsection{Open Areas and Limitations}

While the proposed conceptual framework provides a structured approach to ascertainable fairness, several challenges remain:
\begin{itemize}
    \item Fairness Metrics Standardization: Current fairness measures lack a unified standard, leading to discrepancies between ADM providers and end-user expectations. Further research is needed to harmonize metrics in different applications.
    \item User Literacy: Fairness assessments and contestability mechanisms should be accessible to non-experts. Future work should focus on developing user-friendly tools that facilitate fairness verification without requiring deep technical knowledge.
    \item Implementation of the components: Components 1 and 2 of the framework can be implemented with current tools but need significant optimization to benefit end-users; ideally, they should be implemented by regulatory entities that develop standardized tools for fairness verification, similar to sandboxes for ADM system testing and validation. Component 3 is still unexplored to support users in formulating contestation requests and receiving appropriate justifications. With respect to component 4, effective auditing mechanisms must be standardized and integrated into the governance of ADM. Further efforts should define clear protocols for auditing fairness claims and handling disputes.
    \item Human-In-The-Loop Considerations: This framework primarily addresses fully automated decision-making. Future research should adapt it to hybrid systems where human oversight interacts with ADM systems, ensuring fairness in both automated and human-influenced decisions.
\end{itemize}

The proposed framework for ascertainable fairness is congruent with the requirements delineated in the European Union's AI Act. The framework's focus on user empowerment and the verification of fairness directly supports the AI Act's stipulations for high-risk AI systems, specifically in terms of transparency, human oversight, record-keeping, and accountability. The notion of ascertainable fairness, in particular, strengthens the regulatory emphasis on the protection of fundamental rights. Furthermore, the framework's focus on regulatory entities and auditing mechanisms conforms to the provisions for notified bodies as outlined in the AI Act. However, more research is needed to achieve alignment of implementation details with the specific technical requirements and conformity assessments mandated by the AI Act. While the framework is aligned with current regulatory requirements, the dynamic nature of AI regulation necessitates additional mechanisms to ensure continuous compliance with emerging regulatory requirements.

\section{Conclusions}
\label{sec_8_conc}

This paper introduced a novel framework designed to operationalize the proposed concept of ascertainable fairness. Unlike traditional approaches that focus on providing tools and metrics for practitioners to ensure fairness during the development of ADM systems, our framework shifts the focus towards empowering citizens to directly engage with and ascertain the fairness of the decisions that affect them. In doing so, we enable individuals to materialize their epistemic right to ascertain fairness.

The conceptual framework for ascertain fairness introduces a set of components (fairness of predictions, fairness of recourse, contestation mechanism, and audit mechanism) aimed at empowering users to authenticate fairness based on their personal and group identity, identify possible sources of discrimination, and acquire justifications through contestation. This framework, along with the defined requirements that ADM systems must fulfill to make it possible, allows users to understand, contest, verify, or change the decisions made by ADM systems. Each tool supports users in actively engaging with fairness, from understanding how decisions were made to having the opportunity to challenge those decisions and, if necessary, escalate their concerns to independent audits. Furthermore, this research advocates for the required inclusion of mechanisms that allow for contestability in ADM systems.

The framework marks a shift in the AI fairness landscape by emphasizing procedural fairness from the end-users' point of view and moving beyond the technical, practitioner-centered approaches commonly found in the literature. This proposal places fairness verification in the hands of those directly affected by algorithmic decisions, allowing them to actively participate in ensuring that they are treated fairly.

In addition to its practical benefits for end-users, the proposed framework serves as a guide for policymakers by highlighting the need for clear mechanisms that allow individuals to report and address unfair treatment and the importance of elucidating the concepts of fairness that can be valid and applicable to different contexts. It is also useful for organizations that develop and deploy ADM systems and developers who create them. It shows how people can be empowered to assess and challenge the fairness of AI decisions, helping to ensure that legal and procedural safeguards are in place to monitor, audit, and rectify discrimination, thus strengthening the accountability and reliability of ADM systems.

This work bridges the procedural and substantive dimensions of fairness and ensures that fairness is not only a technical property of ADM systems, but a right that individuals can ascertain and uphold in practice. Allowing end-users to find an answer for \textit{Am I being treated fairly?} employing a systematically organized framework of tools and processes improves transparency, thus increasing the trustworthiness of ADM systems.

\bibliography{main}

\end{document}